\documentclass[pre,aps,preprint,showpacs,preprintnumbers,amsmath,amssymb,secnumroman,eqsecnum]{revtex4}

\usepackage{graphicx}
\usepackage{dcolumn}
\usepackage{bm}

\newcommand{\beq}{\begin{equation}}
\newcommand{\eeq}{\end{equation}}
\newcommand{\beqa}{\begin{eqnarray}}
\newcommand{\eeqa}{\end{eqnarray}}
\newcommand{\vc}[1]{\mbox{\boldmath $#1$}}

\newcommand{\vol}[1]{{\bf #1}}

\newcommand{\du}[1]{{\bf\sf #1}}

\begin{document}


\title{Swimming of a linear chain with a cargo in an incompressible viscous fluid with inertia}

\author{B. U. Felderhof}

 \email{ufelder@physik.rwth-aachen.de}
\affiliation{Institut f\"ur Theorie der Statistischen Physik \\ RWTH Aachen University\\
Templergraben 55\\52056 Aachen\\ Germany\\
}%

\date{\today}

\begin{abstract}
An approximation to the added mass matrix of an assembly of spheres is constructed on the basis of potential flow theory for situations where one sphere is much larger than the others. In the approximation the flow potential near a small sphere is assumed to be dipolar, but near the large sphere it involves all higher order multipoles. The analysis is based on an exact result for the potential of a magnetic dipole in the presence of a superconducting sphere. Subsequently, the approximate added mass hydrodynamic interactions are used in a calculation of the swimming velocity and rate of dissipation of linear chain structures consisting of a number of small spheres and a single large one, with account also of frictional hydrodynamic interactions. The results derived for periodic swimming on the basis of a kinematic approach are compared with bilinear theory, valid for small amplitude of stroke, and with the numerical solution of the approximate equations of motion. The calculations cover the whole range of scale number between the friction-dominated Stokes limit and the inertia-dominated regime.

\end{abstract}

\maketitle
\section{\label{I}Introduction}

In earlier work we have studied the swimming of linear chain structures, consisting of a number of small spheres and a single large one, in the Stokes limit where inertia of the spheres and the fluid is neglected \cite{1}. A model of this kind with two little spheres pushing a big one was first studied by Golestanian \cite{2}. It is of interest to investigate the effects of inertia on the swimming of bodies immersed in an incompressible viscous fluid. Many microorganisms are so small that the Stokes limit provides a valid approximation, but for larger bodies inertia becomes important. The corresponding reactive effects are caused by the dependence of the added mass of the whole structure on its instantaneous configuration.

In models where the structure is approximated as an assembly of spheres the added mass hydrodynamic interactions are manifested as a dependence of the mass matrix on the instantaneous positions of centers. In principle the dependence can be found from the theory of potential flow. Elsewhere we have used a dipole approximation to calculate the mass matrix of an assembly of spheres \cite{3}\cite{4}. The approximation is useful when all spheres are far apart, but fails for configurations where some small spheres are close to a large one. For such configurations an approximate method of calculation of the Stokes friction matrix was proposed by Ekiel-Je\.zewska and Felderhof \cite{5}. In the following we derive a similar approximation to the mass matrix on the basis of an exact expression for the field of a magnetic point dipole in the presence of a superconducting sphere, as derived by Palaniappan \cite{6}. The corresponding scalar potential can be used in hydrodynamics if account is taken of convection of the spheres by the fluid. In Sec. II of this article we present the derivation of the approximate mass matrix in some detail.

Subsequently we apply the approximation in a study of the longitudinal swimming of linear chain structures with a single large sphere and a number of small ones. As an example, we calculate in Sec. III the mean swimming velocity and power for swimming of a linear chain of three small spheres and one large one on the basis of kinematics, taking into account both friction and added mass effects \cite{7}. We assume that the stroke is proportional to the one that is optimal at small amplitude in the Stokes limit, as calculated earlier \cite{1}.

For a linear chain of three identical spheres the optimal stroke, as derived from bilinear theory valid to second order in the amplitude of swimming, is nearly independent of the scale number which characterizes the effect of fluid inertia \cite{4}. In Sec. IV we find that the same is true for chains of three small spheres and an additional large one. We show also that the bilinear theory provides a good approximation to the results derived from the kinematic approach of Sec. III.

In Sec. V we calculate the corresponding actuating forces for a chain with harmonic elastic interactions and compare with the swimming of the chain with application of a cargo constraint, implying that the actuating force on the large sphere vanishes. We use bilinear theory to optimize the actuating forces subject to the cargo constraint. It turns out that the latter does not lead to a significant decrease of swimming efficiency compared to the optimum value without the constraint. Finally, we study transient effects in swimming by solving the approximate equations of motion for the spheres numerically, for a chain subject to actuating forces with the cargo constraint, starting from the rest situation.

\section{\label{II}Added mass in small particle approximation}

As a preliminary we study in this section added mass hydrodynamic interactions for situations where one sphere is much larger than the others. We consider an $N+1$-body system consisting of a sphere of radius $b$, labeled $0$, and $N$ much smaller particles of radii $a_1,...,a_N$, all immersed in a viscous incompressible fluid of shear viscosity $\eta$ and mass density $\rho$. The fluid is of infinite extent in all directions. The whole system is considered to be at rest for $t<0$. At time $t=0$ the system is caused to move by instantaneous forces of short duration
\begin{equation}
\label{2.1}\vc{E}_j(t)=\vc{S}_j\delta(t)\qquad j=0,1,...,N,
\end{equation}
where $\vc{S}_j$ is the impulse imparted to the sphere or particle. In order to satisfy the kinematic boundary condition that sphere and particles are impenetrable, the fluid moves with a flow velocity $\vc{v}(\vc{r})$ satisfying Laplace's equation,
\begin{equation}
\label{2.2}\vc{v}=-\nabla\phi,\qquad\nabla^2\phi=0.
\end{equation}
The hydrodynamic interactions are embodied in the $(3N+3)\times(3N+3)$  inverse mass matrix $\du{w}$ which relates the translational velocities $(\vc{U}_0,...,\vc{U}_N)$ to the $N+1$ impulses $(\vc{S}_0,...,\vc{S}_N)$ imparted to the bodies via \cite{4}
 \begin{equation}
\label{2.3}\vc{U}_j=\sum^N_{k=0}\du{w}_{jk}\cdot\vc{S}_k,\qquad (j=0,...,N),
\end{equation}
 where each tensor $\du{w}_{jk}$ depends on the positions $(\vc{R}_0,...,\vc{R}_N)$ of all centers. Correspondingly the mass matrix $\du{m}=\du{w}^{-1}$ also depends on all center positions. We shall derive approximate simplified expressions for the inverse mass tensors, based on an approximation to the irrotational flow pattern and the corresponding convective effects. In the approximation the element $\du{w}_{00}$ is independent of position, elements $\du{w}_{0j}$ and $\du{w}_{j0}$ with $j>0$ depend only on $\vc{R}_j-\vc{R}_0$ via a pair hydrodynamic interaction, and elements $\du{w}_{jk}$, with $j>0,k>0$ depend only on $\vc{R}_j-\vc{R}_0$ and $\vc{R}_k-\vc{R}_0$ via a three-body hydrodynamic interaction. In order to derive the expression for the pair hydrodynamic interactions $\du{w}_{0j}$ and $\du{w}_{j0}$ it suffices to consider the sphere and a single small particle. In order to derive the three-body hydrodynamic interaction we must consider the sphere and two small particles.

We use Cartesian coordinates with origin at the center of the large sphere. We recall first that in the absence of particles, the sphere made to move with translational velocity $\vc{U}_0$ generates a potential \cite{8}
\begin{equation}
\label{2.4}\phi(\vc{r})=\frac{1}{2}\;b^3\frac{\vc{r}}{r^3}\cdot\vc{U}_0,\qquad r>b,
\end{equation}
corresponding to the dipole moment
 \begin{equation}
\label{2.5}\vc{q}_0=\frac{1}{2}\;b^3\vc{U}_0.
\end{equation}
The corresponding Poisson flow pattern $\vc{v}^P=-\nabla\phi$ is given by
\begin{equation}
\label{2.6}\vc{v}^P(\vc{r})=\du{F}_0(\vc{r})\cdot\vc{q}_0,\qquad \du{F}_0(\vc{r})=\frac{-\du{I}+3\hat{\vc{r}}\hat{\vc{r}}}{r^3},
\end{equation}
where $\du{F}_0(\vc{r})$ is the dipolar tensor with unit tensor $\du{I}$.
The dipole moment is related to the impulse $\vc{S}_0$ by
\begin{equation}
\label{2.7}\vc{q}_0=\beta_0\vc{S}_0,\qquad\beta_0=\frac{b^3}{2m^*_0},
\end{equation}
where $m^*_0$ is the effective mass
\begin{equation}
\label{2.8}m^*_0=m_0+\frac{1}{2}m_{f0},\qquad m_{f0}=\frac{4\pi}{3}\rho b^3.
\end{equation}
Here $m_{f0}$ is the mass of displaced fluid and $\frac{1}{2}m_{f0}$ is the added mass.

Next we consider a sphere of radius $b$ and a single small particle, both immersed in the
fluid. The pair hydrodynamic interaction between particle and sphere is embodied in the six-dimensional inverse mass matrix relating the sphere velocity $\vc{U}_0$ and the particle velocity $\vc{U}_1$ to the impulses $\vc{S}_0$ and $\vc{S_1}$ imparted to the sphere and particle according to
\begin{eqnarray}
\label{2.9}\vc{U}_0&=&\du{w}_{00}\cdot\vc{S}_0+\du{w}_{01}\cdot\vc{S}_1,\nonumber\\
\vc{U}_1&=&\du{w}_{10}\cdot\vc{S}_0+\du{w}_{11}\cdot\vc{S}_1.
\end{eqnarray}
We consider first the situation where an impulse $\vc{S}_0$ is imparted to the sphere, but $\vc{S}_1=0$.
In our approximation the particle is subjected to the dipolar flow pattern generated by $\vc{S}_0$, but its reaction is neglected, except for its convective motion. Hence the parts $\du{w}_{00}$ and $\du{w}_{10}$ of the inverse mass matrix are given simply by
 \begin{equation}
\label{2.10}\du{w}_{00}=\frac{1}{m^*_0}\;\du{I},\qquad\du{w}_{10}=\gamma_1\beta_0\du{F}_0(\vc{r}_1),
\end{equation}
with convective coefficient
 \begin{equation}
\label{2.11}\gamma_1=\frac{3m_{f1}}{2m^*_1}=4\pi\rho\beta_1,
\end{equation}
and relative vector $\vc{r}_1=\vc{R}_1-\vc{R}_0$. The convective coefficient with $m_{f1}=(4\pi/3)\rho a_1^3$ and $m_1^*=m_1+\frac{1}{2}m_{f1}$ follows from linear response theory \cite{9}. The value is consistent with the derivations of Landau and Lifshitz \cite{10} and of Batchelor \cite{11}.

In order to derive an approximate expression for the part $\du{w}_{01}$ of the inverse mass matrix in Eq. (2.9) we consider a flow situation where an impulse $\vc{S}_1$ is applied to the particle, but the sphere is convected such that it exerts no force on the fluid. In the absence of the sphere and in infinite fluid the impulse $\vc{S}_1$ would generate a dipolar flow $\du{F}_0(\vc{r}-\vc{r}_1)\cdot\vc{q}_1$ with dipole moment $\vc{q}_1=\beta_1\vc{S}_1$. We regard this as an unperturbed flow acting on the freely moving sphere. The resulting sphere velocity is \cite{9}
 \begin{equation}
\label{2.12}\vc{U}_{01}=\gamma_0\overline{\du{F}_0(\vc{r}-\vc{r}_1)}^V\cdot\vc{q}_1,
\end{equation}
where the overline indicates an average over the volume of the sphere. Here
 \begin{equation}
\label{2.13}\gamma_0=\frac{3m_{f0}}{2m^*_0}=4\pi\rho\beta_0,
\end{equation}
as in Eq. (2.11). The average in Eq. (2.12) corresponds to the field at $\vc{r}_1$ due to a uniformly polarized sphere. This is dipolar in $\vc{r}_1$, so that we have
 \begin{equation}
\label{2.14}\vc{U}_{01}=\gamma_0\du{F}_0(\vc{r}_1)\cdot\vc{q}_1,
\end{equation}
corresponding to
\begin{equation}
\label{2.15}\du{w}_{01}=\gamma_0\beta_1\du{F}_0(\vc{r}_1).
\end{equation}
Comparing with Eq. (2.10) we see that $\du{w}_{01}=\du{w}_{10}$, since $\gamma_0\beta_1=\gamma_1\beta_0$. There is symmetry, as expected on general grounds.

In order to derive an approximation to the part $\du{w}_{11}$ of the inverse mass matrix we consider the flow generated by the sphere when acted upon by the dipolar flow due to the particle. First we recall an important result derived by Palaniappan \cite{6} for the field of a magnetic dipole near a superconducting sphere. For a fixed sphere centered at the origin and with a point dipole $\vc{q}_1$ acting at $\vc{r}_1$ he derived the scalar potential \cite{6}
\begin{equation}
\label{2.16}\Phi(\vc{r},\vc{r}_1)=\Phi_0(\vc{r}-\vc{r}_1)+\Phi_R(\vc{r},\vc{r}_1),
\end{equation}
where $\Phi_0(\vc{r}-\vc{r}_1)$ is the potential for infinite fluid in the absence of the sphere,
\begin{equation}
\label{2.17}\Phi_0(\vc{r}-\vc{r}_1)=\frac{(\vc{r}-\vc{r}_1)\cdot\vc{q}_1}{|\vc{r}-\vc{r}_1|^3},
\end{equation}
and $\Phi_R(\vc{r},\vc{r}_1)$ is the reflection potential
\begin{eqnarray}
\label{2.18}\Phi_R(\vc{r},\vc{r}_1)&=&\frac{b^3}{r_1^3}\;\frac{(\vc{r}-\overline{\vc{r}_1})\cdot(\vc{q}_{1,\perp}-\vc{q}_{1,\parallel})}{\overline{d}^3}\nonumber\\
&-&\frac{1}{br_1[r^2-(\vc{r}\cdot\vc{r}_1/r_1)^2]}\bigg(r-\frac{\vc{r}\cdot(\vc{r}-\overline{\vc{r}_1})}{\overline{d}}\bigg)\vc{r}\cdot\vc{q}_{1,\perp},
\end{eqnarray}
where the image point $\overline{\vc{r}_1}$ is defined by
\begin{equation}
\label{2.19}\overline{\vc{r}_1}=\frac{b^2}{r_1^2}\vc{r}_1,
\end{equation}
and $\overline{d}$ is the distance from the field point $\vc{r}$ to the image point,
 \begin{equation}
\label{2.20}\overline{d}=|\vc{r}-\overline{\vc{r}_1}|.
\end{equation}
Furthermore, the vector
 \begin{equation}
\label{2.21}\vc{q}_{1,\parallel}=\frac{\vc{r}_1\vc{r}_1}{r_1^2}\cdot\vc{q}_1
\end{equation}
is the part of the dipole moment parallel to $\vc{r}_1$ and $\vc{q}_{1,\perp}=\vc{q}_1-\vc{q}_{1,\parallel}$ is the part perpendicular to $\vc{r}_1$. The potential $\Phi$ satisfies Laplace's equation $\nabla^2\Phi=0$, as well as the boundary condition
 \begin{equation}
\label{2.22}\frac{\partial\Phi}{\partial r}=0\qquad\mathrm{on}\;r=b.
\end{equation}
Hence the corresponding flow $\vc{v}=-\nabla\Phi$ is tangential to the spherical surface. The flow can be expressed as
 \begin{equation}
\label{2.23}\vc{v}_1(\vc{r})=-\nabla\Phi(\vc{r},\vc{r}_1)=\du{F}(\vc{r},\vc{r}_1)\cdot\vc{q}_1,
\end{equation}
with Green tensor $\du{F}(\vc{r},\vc{r}_1)$. The Green tensor can be decomposed as
 \begin{equation}
\label{2.24}\du{F}(\vc{r},\vc{r}_1)=\du{F}_0(\vc{r}-\vc{r}_1)+\du{F}_R(\vc{r},\vc{r}_1),
\end{equation}
where $\du{F}_R(\vc{r},\vc{r}_1)$ is the reflection tensor. The tensors have the symmetry
 \begin{equation}
\label{2.25}\du{F}_0(\vc{r}-\vc{r}_1)=\du{F}_0(\vc{r}_1-\vc{r})^T,\qquad
\du{F}_R(\vc{r},\vc{r}_1)=\du{F}_R(\vc{r}_1,\vc{r})^T,
\end{equation}
where the superscript $T$ indicates the transpose.

In order to obtain the flow generated by $\vc{q}_1$ in the presence of the freely moving sphere we must add the dipolar flow corresponding to the sphere velocity $\vc{U}_{01}$ given by Eq. (2.14). The total flow can be expressed as
 \begin{equation}
\label{2.26}\hat{\vc{v}}_1(\vc{r})=\hat{\du{F}}(\vc{r},\vc{r}_1)\cdot\vc{q}_1,
\end{equation}
with Green tensor
 \begin{equation}
\label{2.27}\hat{\du{F}}(\vc{r},\vc{r}_1)=\du{F}(\vc{r},\vc{r}_1)+\du{V}(\vc{r},\vc{r}_1)
\end{equation}
with tensor function $\du{V}(\vc{r},\vc{r}_1)$ given by
 \begin{equation}
\label{2.28}\du{V}(\vc{r},\vc{r}_1)=\frac{1}{2}\;b^3\gamma_0\du{F}_0(\vc{r})\cdot\du{F}_0(\vc{r}_1).
\end{equation}
 The tensor has the symmetry
 \begin{equation}
\label{2.29}\du{V}(\vc{r},\vc{r}_1)=\du{V}(\vc{r}_1,\vc{r})^T.
\end{equation}

The total image dipole of the fixed sphere is
\begin{equation}
\label{2.30}\overline{\vc{q}_1}=-\frac{1}{2}b^3\du{F}_0(\vc{r}_1)\cdot\vc{q}_1,
\end{equation}
as one sees from the long range behavior of the reflection potential $\Phi_R$ in Eq. (2.18). For a neutrally buoyant sphere $\gamma_0=1$, and then this is precisely canceled by the dipole corresponding to the flow caused by sphere velocity $\vc{U}_{01}$. More generally, the tensor $\hat{\du{F}}_R(\vc{r},\vc{r}_1)$ tends to
\begin{equation}
\label{2.31}\hat{\du{F}}_R(\vc{r},\vc{r}_1)\approx(\gamma_0-1)\frac{1}{2}\;b^3\du{F}_0(\vc{r})\cdot\du{F}_0(\vc{r}_1),\qquad (r>>b)
\end{equation}
at large distance from sphere and particle.

We derive an approximate expression for the inverse mass tensor $\du{w}_{11}$ by considering the reflected part of the modified flow $\hat{\vc{v}}_1(\vc{r})$ at the point $\vc{r}=\vc{r}_1$. This yields
\begin{equation}
\label{2.32}\du{w}_{11}=\frac{1}{m^*_1}\;\du{I}+\beta_1\gamma_1\hat{\du{F}}_R(\vc{r}_1,\vc{r}_1).
\end{equation}

From Eqs. (2.18) and (2.27) we find the explicit expression
\begin{eqnarray}
\label{2.33}\hat{\du{F}}_R(\vc{r}_1,\vc{r}_1)&=&\frac{-2b^3}{(r_1^2-b^2)^3}\;\hat{\vc{r}}_1\hat{\vc{r}}_1
-\frac{b^2+r_1^2}{2r_1^2}\frac{b^3}{(r_1^2-b^2)^3}\;(\du{I}-\hat{\vc{r}}_1\hat{\vc{r}}_1)\nonumber\\
&+&\frac{2b^3\gamma_0}{r_1^6}\;\hat{\vc{r}}_1\hat{\vc{r}}_1
+\frac{b^3\gamma_0}{2r_1^6}\;(\du{I}-\hat{\vc{r}}_1\hat{\vc{r}}_1).
\end{eqnarray}
The derivation of the first two terms is best performed in spherical coordinates.
The last two terms depend on the mass of the sphere via $\gamma_0$. The expression satisfies Eq. (2.31), since
\begin{equation}
\label{2.34}\du{F}_0(\vc{r}_1)\cdot\du{F}_0(\vc{r}_1)=\frac{4}{r_1^6}\;\hat{\vc{r}}_1\hat{\vc{r}}_1+\frac{1}{r_1^6}\;(\du{I}-\hat{\vc{r}}_1\hat{\vc{r}}_1).
\end{equation}
Note that $\hat{\du{F}}_R(\vc{r}_1,\vc{r}_1)$ becomes singular at short distance $r_1-b$ and shows a rapid decay at large distance. The tensor is obviously symmetric. In our approximation it involves only hydrodynamic interactions between particle 1 and the big sphere. It can be shown that the above results are consistent with the dipole approximation \cite{3}\cite{4} and with the calculation of the kinetic energy of irrotational flow for two moving spheres quoted by Lamb \cite{12}.

We can use the same formalism to derive the approximate expression for the inverse mass tensor $\du{w}_{jk}$, with $j>0,k>0$. It suffices to consider a sphere and two small particles located at $\vc{R}_1$ and $\vc{R}_2$. The linear relation between velocities and impulses becomes
\begin{eqnarray}
\label{2.35}\vc{U}_0&=&\du{w}_{00}\cdot\vc{S}_0+\du{w}_{01}\cdot\vc{S}_1+\du{w}_{02}\cdot\vc{S}_2,\nonumber\\
\vc{U}_1&=&\du{w}_{10}\cdot\vc{S}_0+\du{w}_{11}\cdot\vc{S}_1+\du{w}_{12}\cdot\vc{S}_2,\nonumber\\
\vc{U}_2&=&\du{w}_{20}\cdot\vc{S}_0+\du{w}_{21}\cdot\vc{S}_1+\du{w}_{22}\cdot\vc{S}_2.
\end{eqnarray}
Most parts of the inverse mass matrix are given by expressions derived above. Only the parts $\du{w}_{12}$ and $\du{w}_{21}$ require further consideration.
In generalization of Eq. (2.32) we have
\begin{equation}
\label{2.36}\du{w}_{12}=\gamma_1\beta_2\hat{\du{F}}(\vc{r}_1,\vc{r}_2),\qquad\du{w}_{21}=\gamma_2\beta_1\hat{\du{F}}(\vc{r}_2,\vc{r}_1),
\end{equation}
with $\vc{r}_1=\vc{R}_1-\vc{R_0}$ and $\vc{r}_2=\vc{R}_2-\vc{R_0}$.
The symmetry relation
\begin{equation}
\label{2.37}\du{w}_{21}=\du{w}^T_{12}
\end{equation}
is satisfied, as follows from Eqs. (2.25) and (2.29), in agreement with general arguments.

The expression for $\du{w}_{12}$ given by Eq. (2.36) is complicated, but it simplifies for configurations for which $\vc{r}_1$ and $\vc{r}_2$ are collinear with the origin, so that $\hat{\vc{r}}_1=\hat{\vc{r}}_2$. For such configurations we find
\begin{eqnarray}
\label{2.38}\du{w}_{12}&=&\frac{\gamma_1\gamma_2}{4\pi\rho}\bigg[\du{F}_0(\vc{r}_1-\vc{r}_2)-\frac{2b^3}{(r_1r_2-b^2)^3}\;\hat{\vc{r}}_1\hat{\vc{r}}_1\nonumber\\
&-&\frac{b^2+r_1r_2}{2r_1r_2}\frac{b^3}{(r_1r_2-b^2)^3}\;(\du{I}-\hat{\vc{r}}_1\hat{\vc{r}}_1)+\frac{b^3\gamma_0}{2r_1^3r_2^3}\;(\du{I}+3\hat{\vc{r}}_1\hat{\vc{r}}_1)\bigg].\nonumber\\
\end{eqnarray}
The first term is the pair hydrodynamic interaction between two small particles.
The last three terms depend on both $\vc{r}_1=\vc{R}_1-\vc{R}_0$ and $\vc{r}_2=\vc{R}_2-\vc{R}_0$, and therefore represent a three-body hydrodynamic interaction. Note that the interaction becomes singular as both $r_1$ and $r_2$ tend to the sphere radius $b$. It decays as $1/(r_1r_2)^3$ as both $r_1$ and $r_2$ tend to infinity.

The expressions derived above can be used as approximations in the many-body inverse mass matrix in Eq. (2.3). If we regard $a$ as a typical small particle radius, $d$ as a typical distance between small particles, and $h$ as the minimum separation distance of the center of a small particle from the surface of the sphere, then the ratios $a/b$, $a/d$ and $a/h$ may be regarded as small parameters. It follows from a consideration of the multipole expansion of the exact mass matrix that our expressions represent the first few terms in a systematic expansion in powers of the small parameters. We call our expressions the small particle approximation to the inverse mass matrix.

\section{\label{III}Kinematic swimming}

As an application of the added mass hydrodynamic interactions derived in the preceding section we consider swimming of a linear chain structure consisting of a big sphere of radius $b$ and three small spheres of radius $a$ with periodic motions of the four bodies along the $x$ axis of a Cartesian system of coordinates. In Fig. 1 we show a sketch of the structure. The small spheres have centers at $x_1(t),x_2(t),x_3(t)$ with $x_1<x_2<x_3$ and the center of the big sphere is at $x_4(t)$ with $x_4>x_3$. In this section we use a kinematic approach and prescribe the periodic relative motion $\du{r}(t)=(r_1(t),r_2(t),r_3(t))$ of the four spheres, where $r_1=x_2-x_1,\;r_2=x_3-x_2,\;r_3=x_4-x_3$. For given relative motion of the spheres the asymptotic periodic swimming velocity and the periodic rate of dissipation are determined by the added mass and frictional hydrodynamic interactions. We use a $4\times 4$ mass matrix $\du{m}=\du{w}^{-1}$ as derived in the preceding section, and a $4\times 4$ friction matrix $\vc{\zeta}=\vc{\mu}^{-1}$ as derived in earlier work with Ekiel-Je\.zewska \cite{5}.

The asymptotic periodic swimming velocity $U_{sw}(t)$ can be calculated from the periodic total mass $M(t)$ and total friction coefficient $Z(t)$ by use of an expression derived earlier \cite{7}. From the given relative motion and the calculated swimming velocity the periodic sphere velocities in the laboratory frame can be evaluated. The velocities of the individual spheres are a sum of the swimming velocity and a displacement velocity. The latter is a component of the vector
 \begin{equation}
\label{3.1}\dot{\du{d}}=\du{T}^{-1}
\cdot(0,\frac{d\du{r}}{dt}),
\end{equation}
where $\du{T}$ is the matrix relating center $X$ and relative coordinates $(r_1,r_2,r_3)$ to the Cartesian coordinates $(x_1,x_2,x_3,x_4)$. In the present case this is given explicitly by
\begin{eqnarray}
\label{3.2}\du{T}=\left(\begin{array}{cccc}
\frac{1}{4}&\frac{1}{4}&\frac{1}{4}&\frac{1}{4}
\\-1&1&0&0\\0&-1&1&0\\0&0&-1&1\end{array}\right).
\end{eqnarray}
The periodic rate of dissipation $\mathcal{D}(t)$ then follows from the expression $\mathcal{D}=\du{U}\cdot\vc{\zeta}\cdot\du{U}$, where $\du{U}=U_{sw}\du{u}+\dot{\du{d}}$, with $\du{u}=(1,1,1,1)$, is the four-vector comprising the $x$-components of the sphere velocities. It follows from Eq. (3.1) that $\du{u}\cdot\dot{\du{d}}=0$.

The reduced speed $\hat{U}$ is defined by $|\overline{U}_{sw}|=\omega a\hat{U}$, where the overhead bar indicates the time average over a period $\mathcal{T}=2\pi/\omega$, and the reduced rate of dissipation $\hat{\mathcal{D}}$ is defined by $\overline{\mathcal{D}}=\eta\omega^2a^3\hat{\mathcal{D}}$. For small amplitude swimming we denote the reduced quantities as $\varepsilon^2\hat{U}_2$ and $\varepsilon^2\hat{\mathcal{D}}_2$, where $\varepsilon$ is an amplitude factor and the subscript 2 indicates that the calculation is performed to second order in the amplitude. We study these quantities as functions of the dimensionless scale number $s$ defined by \cite{13}
 \begin{equation}
\label{3.3}s^2=\frac{a^2\omega\rho}{2\eta}.
\end{equation}
For $s>>1$ the swimming is dominated by added mass effects, whereas $s=0$ corresponds to the Stokes limit, dominated by friction. The efficiency of the stroke is defined by
 \begin{equation}
\label{3.4}E_T=\frac{\hat{U}}{\hat{\mathcal{D}}}.
\end{equation}
The time-dependent periodic swimming velocity and rate of dissipation are calculated from an expansion in Fourier series \cite{7}.

In earlier work we have studied the swimming of a collinear chain in the Stokes limit \cite{1} and determined the optimal stroke for small amplitude swimming of a chain of harmonically linked spheres with equilibrium relative distances $\du{r}_0=(d,d,b+d)$. The optimal stroke is independent of the strength of direct interactions, since it is determined solely by kinematics. The optimal stroke follows from an eigenvalue problem with a matrix $\du{B}$ determining $\hat{U}_2=\frac{1}{2}(\vc{\xi}^c|\du{B}|\vc{\xi}^c)$ and a matrix $\du{A}$ determining $\hat{\mathcal{D}}_2=\frac{1}{2}(\vc{\xi}^c|\du{A}|\vc{\xi}^c)$, where $|\vc{\xi}^c)$ is the complex vector of relative displacements. The maximum eigenvalue $\lambda_{max}$ of the eigenvalue problem $\du{B}|\vc{\xi}^c)=\lambda\du{A}|\vc{\xi}^c)$ corresponds to the optimal ratio $\hat{U}_2/\hat{\mathcal{D}}_2$. For the case $d=5a,\;b=10a$ we found $\lambda_{max}=930\times 10^{-7}$ with corresponding complex eigenvector
 \begin{equation}
\label{3.5}\hat{\vc{\xi}}^c_0=(0.720,-0.121+0.598i,-0.311-0.112i).
\end{equation}
The vector is normalized to unity. The corresponding reduced speed and power are
 \begin{equation}
\label{3.6}\hat{U}_2=0.000792,\qquad\hat{\mathcal{D}}_2=8.523,\qquad (s=0).
\end{equation}
The eigenvector determines the optimal stroke in the Stokes limit. The relative motion during the optimal stroke is given by
 \begin{equation}
\label{3.7}\du{r}(t)=\du{r}_0+\varepsilon a\;\mathrm{Re}\;[\hat{\vc{\xi}}^c_0e^{-i\omega t}].
\end{equation}
with amplitude factor $\varepsilon$. In Fig. 2 we show the motion of the four spheres for $\varepsilon=2$ in the center system, where the geometrical center $X$ of the four spheres is at rest. For comparison, the maximum eigenvalue of the 3-chain without the big sphere in the Stokes limit is $\lambda_{max}=949\times 10^{-6}$, as given by Eq. (6.5) in Ref. 14. The corresponding values of reduced speed and power are $\hat{U}_2=0.00639$ and $\hat{\mathcal{D}}_2=6.732$.

For the system with inertia we use the same stroke, given by Eq. (3.5). We have found for the chain with three identical spheres \cite{4} that the optimal stroke at small amplitude varies little as a function of scale number $s$, and we shall show in Sec. IV that the same is true for the present system.

We assume that the four spheres are neutrally buoyant. In particular we then find numerically for $s=1$ and $\varepsilon=2$
 \begin{equation}
\label{3.8}\hat{U}=0.00328,\qquad\hat{\mathcal{D}}=34.381,\qquad (s=1,\;\varepsilon=2),
\end{equation}
corresponding to efficiency $E_T=955\times 10^{-7}$. For $s^2=10$ and $\varepsilon=2$ we find
 \begin{equation}
\label{3.9}\hat{U}=0.00317,\qquad\hat{\mathcal{D}}=34.381,\qquad (s^2=10,\;\varepsilon=2),
\end{equation}
corresponding to efficiency $E_T=907\times 10^{-7}$. For comparison, in the Stokes limit and to second order in the amplitude
 \begin{equation}
\label{3.10}\varepsilon^2\hat{U}_2=0.00317,\qquad\varepsilon^2\hat{\mathcal{D}}_2=34.092,\qquad (s=0,\;\varepsilon=2),
\end{equation}
corresponding to efficiency $E_T=\lambda_{max}=930\times 10^{-7}$.

The Fourier coefficients of the swimming velocity decrease rapidly with increasing order. For the above case with $s=1,\;\varepsilon=2$ we find for the first five absolute ratios
\begin{equation}
\label{3.11}\{1,\bigg|\frac{U_{sw,1}}{\overline{U_{sw}}}\bigg|,\bigg|\frac{U_{sw,2}}{\overline{U_{sw}}}\bigg|,
\bigg|\frac{U_{sw,3}}{\overline{U_{sw}}}\bigg|,\bigg|\frac{U_{sw,4}}{\overline{U_{sw}}}\bigg|\}=\{1,73.80,10\times 10^{-4},4\times 10^{-4},10^{-5}\}.
\end{equation}
For the case with $s^2=10,\;\varepsilon=2$ we find
\begin{equation}
\label{3.12}\{1,\bigg|\frac{U_{sw,1}}{\overline{U_{sw}}}\bigg|,\bigg|\frac{U_{sw,2}}{\overline{U_{sw}}}\bigg|,
\bigg|\frac{U_{sw,3}}{\overline{U_{sw}}}\bigg|,\bigg|\frac{U_{sw,4}}{\overline{U_{sw}}}\bigg|\}=\{1,77.76,5\times 10^{-4},4\times 10^{-4},1\times 10^{-4}\}.
\end{equation}
In both cases the swimming velocity is essentially just the sum of the mean and the first harmonic, and the amplitude of the first harmonic is larger than the mean.

\section{\label{IV}Bilinear theory}

Next we study small amplitude swimming in a fluid with inertia. We consider $N$ collinear spheres with dynamics dominated by frictional and added mass hydrodynamic interactions. The bilinear theory yields results for mean swimming velocity and power valid to second order in the amplitude of the stroke. These quantities are evaluated from matrices $\du{B}$ and $\du{A}$ operating in the space of relative displacements as
 \begin{equation}
\label{4.1}\overline{U^{(2)}}=\frac{1}{2}\omega a (\vc{\xi}^c|\du{B}|\vc{\xi}^c),\qquad\overline{\mathcal{D}^{(2)}}=\frac{1}{2}\eta\omega^2a^3 (\vc{\xi}^c|\du{A}|\vc{\xi}^c),
\end{equation}
with a typical sphere radius $a$. The $(N-1)$-dimensional vector $\vc{\xi}^c=\vc{\xi}/a$ and the matrices $\du{B}$ and $\du{A}$ have dimensionless elements. The expressions given earlier \cite{4} for the matrices need correction, as detailed below. The numerical calculations in the earlier work \cite{4} were performed with the correct expressions.

It was shown earlier \cite{4} that for harmonic displacements $\du{d}(t)=\mathrm{Re}\;[\du{d}_\omega\exp(-i\omega t)]$ the second order mean swimming velocity is given by
 \begin{equation}
\label{4.2}\overline{U^{(2)}}=\frac{1}{2}\mathrm{Re}\;[i\omega\du{d}^*_\omega\cdot\du{V}(\omega)\big|_0\cdot\du{d}_\omega],
\end{equation}
with frequency-dependent matrix $\du{V}(\omega)$ given by
\begin{equation}
\label{4.3}\du{V}(\omega)=\mu\breve{\du{D}}(\omega),
\end{equation}
with mobility $\mu=\du{u}\cdot\vc{\mu}\cdot\du{u}$, where $\du{u}=(1,...,1)$, and reduced derivative friction matrix
\begin{equation}
\label{4.4}\breve{\du{D}}(\omega)=\du{D}- Y(\omega)\du{g}\;\du{f}(\omega),
\end{equation}
with admittance
\begin{equation}
\label{4.5}Y(\omega)=\big[-i\omega M+Z\big]^{-1},
\end{equation}
and vectors
\begin{equation}
\label{4.6}\du{g}=\du{D}\cdot\du{u},\qquad\du{f}(\omega)=(-i\omega\du{m}+\vc{\zeta})\cdot\du{u}.
\end{equation}
Here $M=\du{u}\cdot\du{m}\cdot\du{u},\;Z=\du{u}\cdot\vc{\zeta}\cdot\du{u}$, and the derivative friction matrix $\du{D}$ is given by
\begin{equation}
\label{4.7}\du{D}=\vc{\nabla}\du{f},\qquad\du{f}=\vc{\zeta}\cdot\du{u},
\end{equation}
with gradient operator $\vc{\nabla}=(\partial/\partial x_1,...,\partial/\partial x_N)$. In Eq. (4.2) the matrix $\du{V}(\omega)$ is calculated in the time-independent equilibrium configuration. In the expression the product $i\omega\du{V}(\omega)$ can be replaced by its hermitian part
\begin{equation}
\label{4.8}[i\omega\du{V}(\omega)]^h=i\omega\du{V}^a,\qquad\du{V}^a=\frac{1}{2}(\du{V}-\du{V}^\dagger).
\end{equation}

Similarly the second order mean rate of dissipation is given by
 \begin{equation}
\label{4.9}\overline{\mathcal{D}^{(2)}}=\frac{1}{2}\omega^2\mathrm{Re}\;[\du{d}^*_\omega\cdot\du{P}(\omega)\big|_0\cdot\du{d}_\omega],
\end{equation}
with the matrix
 \begin{equation}
\label{4.10}\du{P}(\omega)=-i\omega\du{m}+\vc{\zeta}-Y(\omega)\du{f}(\omega)\du{f}(\omega).
\end{equation}
In Eq. (4.9) the matrix $\du{P}(\omega)$ can be replaced by its hermitian part $\du{P}^h=\frac{1}{2}(\du{P}+\du{P}^\dagger)$.

The matrices $\du{V}(\omega)$ and $\du{P}(\omega)$ have the properties
 \begin{eqnarray}
\label{4.11}\du{u}\cdot\du{V}(\omega)&=&0,\qquad\du{V}(\omega)\cdot\du{u}=0,\nonumber\\
\du{u}\cdot\du{P}(\omega)&=&0,\qquad\du{P}(\omega)\cdot\du{u}=0,
\end{eqnarray}
and this allows to reduce the calculation to one in relative space. We define the transformed matrices
 \begin{equation}
\label{4.12}\du{V}_T^a=\du{T}\cdot\du{V}^a\cdot\du{T}^{-1},\qquad\du{P}_T^h=\du{T}\cdot\du{P}^h\cdot\du{T}^{-1},
\end{equation}
where $\du{T}$ is the matrix relating center and relative coordinates in generalization of Eq. (3.2).
The first row and column of the matrices $\du{V}_T^a$ and $\du{P}_T^h$ vanish identically on account of the properties Eq. (4.11). The truncated matrices obtained by erasing the first row and column are denoted as $\hat{\du{V}}_T^a$ and $\hat{\du{P}}_T^h$. Finally, the $(N-1)\times(N-1)$ dimensional matrices $\du{B}$ and $\du{A}$ in Eq. (4.1) are given by
 \begin{equation}
\label{4.13}\du{B}=ia\du{C}_T\cdot\hat{\du{V}}_T^a,\qquad\du{A}=\frac{1}{\eta a}\;\du{C}_T\cdot\hat{\du{P}}_T^h,
\end{equation}
with the matrix
\begin{equation}
\label{4.14}\du{C}_T=[(\du{T}^{-1})^T\cdot\du{T}^{-1}]\;\vc{\hat{}}.
\end{equation}
This $(N-1)\times(N-1)$ dimensional matrix consists of numerical coefficients and is obtained from the corresponding $N\times N$ matrix by truncation, as indicated by the final hat symbol.

In the present context the bilinear theory can be used for two purposes. On the one hand we can study the optimal stroke, as determined from the eigenvalue problem
 \begin{equation}
\label{4.15}\du{B}|\vc{\xi}^c)=\lambda\du{A}|\vc{\xi}^c),
\end{equation}
as the solution with maximum eigenvalue $\lambda_{max}$, and show that for neutrally buoyant spheres this depends only weakly on the scale number $s$, defined in Eq. (3.3). On the other hand, for a calculated or assumed periodic relative motion $\vc{\xi}(t)$ we can obtain a quick estimate of mean swimming velocity and power as functions of the scale number. For example, we can use the stroke given by Eq. (3.5), or the long-time nearly periodic motion found from the numerical solution of the equations of motion with cargo constraint in Sec. V, approximating this by its first harmonic. Furthermore, it is of interest to show that the bilinear theory provides a good approximation over a wide range of amplitudes.

We consider first the eigenvalue problem Eq. (4.15). In Fig. 3 we plot the maximum eigenvalue for $d=5a,\;b=10a$ and neutrally buoyant spheres as a function of $s^2$. The function is nearly constant, but shows a minimum at $s^2=0.0145$. At the minimum $\lambda_{max}=920\times 10^{-7}$. The corresponding eigenvector is
 \begin{equation}
\label{4.16}\hat{\vc{\xi}}^c_0=(0.719,-0.127+0.596i,-0.310-0.122i).
\end{equation}
The eigenvalue tends to $\lambda_{max}=931\times 10^{-7}$ as $s\rightarrow\infty$. The
corresponding eigenvector is
 \begin{equation}
\label{4.17}\hat{\vc{\xi}}^c_0=(0.722,-0.123+0.595i,-0.310-0.114i).
\end{equation}
These values should be compared to Eq. (3.5) corresponding to $s=0$. As can be seen, the eigenvalue and eigenvector hardly vary over the whole range of $s$. This justifies the use of the Stokes eigenvector Eq. (3.5) in the whole range.

Finally we compare the values of mean swimming velocity and power for the bilinear theory in the inertial regime $s>>1$ with those of the exact calculation. The values for the bilinear theory, as calculated for the Stokes eigenvector Eq. (3.5), are
\begin{equation}
\label{4.18}\varepsilon^2\hat{U}_2=0.00320,\qquad\varepsilon^2\hat{\mathcal{D}}_2=34.394,\qquad (s\rightarrow\infty,\;\varepsilon=2),
\end{equation}
corresponding to efficiency $E_T=931\times 10^{-7}$.
These values should be compared with Eq. (3.9). The bilinear theory provides a good estimate even at this large amplitude factor.

 \section{\label{V}Swimming with the cargo constraint}

In the calculations of Sec. III the relative motion of the spheres was prescribed. The asymptotic periodic swimming velocity was derived as a solution of the equation of motion
 \begin{equation}
\label{5.1}\frac{d}{dt}\;(MU)+ZU=\mathcal{I}
\end{equation}
with time-dependent mass $M(t)$ and friction coefficient $Z(t)$ given by
 \begin{equation}
\label{5.2}M=\du{u}\cdot\du{m}\cdot\du{u},\qquad Z=\du{u}\cdot\vc{\zeta}\cdot\du{u},\qquad\du{u}=(1,...,1),
\end{equation}
and impetus $\mathcal{I}(t)$ given by
\begin{equation}
\label{5.3}\mathcal{I}(t)=-\frac{d}{dt}(\du{u}\cdot\du{m}\cdot\dot{\du{d}})-\du{u}\cdot\vc{\zeta}\cdot\dot{\du{d}},
\end{equation}
where $\dot{\du{d}}$, given by Eq. (3.1), is the time-derivative of the displacement vector $\du{d}(t)$. For $N$ collinear spheres with positions $\du{R}(t)=(x_1(t),...,x_N(t))$ and velocities $\du{U}(t)=(U_1(t),...,U_N(t))$ the center velocity is defined as $U(t)=\du{u}\cdot\du{U}(t)/N$. The equation of motion for the center velocity Eq. (5.1) was derived from the equations of motion for the individual spheres, which read \cite{4}
\begin{equation}
\label{5.4}\frac{d\du{R}}{dt}=\du{U},\qquad\frac{d\du{p}}{dt}=-\frac{\partial\mathcal{K}}{\partial\du{R}}-\vc{\zeta}\cdot\du{U}-\frac{\partial V_{int}}{\partial\du{R}}+\du{E},
\end{equation}
with momenta $\du{p}=\du{m}\cdot\du{U}$ and $\mathcal{K}=\frac{1}{2}\du{p}\cdot\du{w}\cdot\du{p}$. The partial derivative $\partial/\partial\du{R}$ is taken at constant momenta $\du{p}$. Furthermore $V_{int}$ is the potential of direct interaction forces. The actuating forces are summarized in $\du{E}(t)=(E_1,...,E_N)$. These oscillate at frequency $\omega$ and sum to zero at any time. The cargo constraint implies that the actuating force on the big sphere vanishes. In our case $E_4(t)$=0, and there are only two independent actuating forces $E_1(t), E_2(t)$, with $E_3(t)=-E_1(t)-E_2(t)$.

We assume the direct interaction forces to be harmonic of the form
\begin{equation}
\label{5.5}-\frac{\partial V_{int}}{\partial\du{R}}=\du{H}\cdot(\du{R}-\du{R_0}),
\end{equation}
with a real and symmetric matrix $\du{H}$ with the property $\du{H}\cdot\du{u}=0$. Specifically we choose
\begin{eqnarray}
\label{5.6}\du{H}=k\left(\begin{array}{cccc}
-1&1&0&0
\\1&-2&1&0\\0&1&-3&2\\0&0&2&-2\end{array}\right)
\end{eqnarray}
with elastic constant $k$. This corresponds to nearest neighbor interactions between the three small spheres and a twice as strong link between the last small sphere and the big sphere. The stiffness of the assembly is characterized by the dimensionless number $\sigma$ defined by $\sigma=k/(\pi\eta a\omega)$. The elastic forces must be chosen such that stability in the numerical calculations is ensured. We choose $s^2=10$ and $\sigma=40$.

For given actuating forces, oscillating at frequency $\omega$ and with the property $\du{u}\cdot\du{E}=0$, we can solve the first order equations of motion,
 \begin{equation}
\label{5.7}\frac{d\du{R}^{(1)}}{dt}=\du{U}^{(1)},\qquad\du{m}^0\cdot\frac{d\du{U}^{(1)}}{dt}=-\vc{\zeta}^0\cdot\du{U}^{(1)}+\du{H}\cdot\du{R}^{(1)}+\du{E},
\end{equation}
as well as the linearized version of Eq. (5.1), to obtain the oscillating displacement $\du{d}(t)$ with the property $\du{u}\cdot\du{d}(t)=0$. The cargo constraint $E_N(t)=0$ limits the class of displacements. The question arises how to select actuating forces, satisfying the cargo constraint, such that the efficiency is maximal.
In the bilinear theory we can optimize by solving the eigenvalue problem in a reduced subspace.

Assuming harmonic time-dependence we find from the linearized version of Eq. (5.1) for the amplitude of the first order swimming velocity \cite{4}
 \begin{equation}
\label{5.8}U^{(1)}_\omega=[-i\omega M^0+Z^0]^{-1}\du{u}\cdot(\omega^2\du{m}^0+i\omega\vc{\zeta}^0)\cdot\du{d}_\omega.
\end{equation}
The first order velocities are $\du{U}^{(1)}_\omega=U^{(1)}_\omega\du{u}-i\omega\du{d}_\omega$.
Using Eq. (3.1) we find from Eq. (5.7) that the relation $\du{u}\cdot\du{E}_{\omega}=0$ is satisfied automatically, and we obtain a linear relation between the amplitude vector in relative space $\vc{\xi}^c_\omega$ and the force amplitude vector $\du{E}_{\omega}$.

In our case the force amplitude vector is assumed to have the form $(E_1,E_2,-E_1-E_2,0)$, and substitution of the relation into $({\vc{\xi}^c}'|\du{B}|\vc{\xi}^c)$ and $({\vc{\xi}^c}'|\du{A}|\vc{\xi}^c)$ leads to a two-dimensional reduction $\du{B}_{EC},\;\du{A}_{EC}$ of the speed matrix and the power matrix, acting in the space of complex vectors $\du{E}_C=(E_1,E_2)$. The eigenvector with maximum eigenvalue of the two-dimensional eigenvalue problem $\du{B}_{EC}|\du{E}_C)=\lambda_{EC}\du{A}_{EC}|\du{E}_C)$ provides the optimal actuating forces satisfying the cargo constraint. The eigenvalue $\lambda_{ECmax}$ equals the maximum efficiency that can be achieved subject to the cargo constraint. The corresponding eigenvector
yields
 \begin{equation}
\label{5.9}\hat{U}_{2EC}=\frac{1}{2}(\du{E}_C|\du{B}_{EC}|\du{E}_C),\qquad\hat{\mathcal{D}}_{2EC}=\frac{1}{2}(\du{E}_C|\du{A}_{EC}|\du{E}_C).
\end{equation}
As before we normalize such that the corresponding vector $\vc{\xi}^c_\omega$ is normalized to unity. In our example we find with amplitude factor $\varepsilon$
 \begin{equation}
\label{5.10}\varepsilon^2\hat{U}_{2EC}=0.00265,\qquad\varepsilon^2\hat{\mathcal{D}}_{2EC}=34.110,\qquad E_T=\lambda_{ECmax}=776\times 10^{-7},\qquad (\varepsilon=2).
\end{equation}
Comparing with the value below Eq. (4.18) we see that the maximum efficiency under the cargo constraint is surprisingly high. It is not much smaller than the optimum efficiency without the constraint.

So far we have considered the asymptotic periodic motion. It is of interest to consider also the transient effects corresponding to the equations of motion Eq. (5.4).
We choose initial conditions corresponding to the rest configuration
 \begin{eqnarray}
\label{5.11}x_1(0)&=&0,\qquad x_2(0)=d,\qquad x_3(0)=2d,\qquad x_4(0)=b+3d,\nonumber\\
p_1(0)&=&0,\qquad p_2(0)=0,\qquad p_3(0)=0\qquad p_4(0)=0.
\end{eqnarray}
The relaxation rate determining the timescale on which the asymptotic periodic swimming velocity is attained is
 \begin{equation}
\label{5.12}\gamma_b=\frac{Z_b}{M_b}=\frac{9\eta}{2\rho b^2}=\frac{9a^2\omega}{4b^2s^2}.
\end{equation}
In our example this equals $\gamma_b=9\omega/4000$, so that many periods are required to attain asymptotic behavior. For a system with cargo constraint we choose the optimum actuating forces, as calculated above. The equations of motion Eq. (5.4) can be solved with the initial conditions Eq. (5.11) and the chosen actuating forces. We do not need the iterative procedure used in previous work \cite{4}. Typically we solve the equations of motion for fifty periods. After fifty periods we can repeat the procedure with the final values of coordinates and momenta as initial conditions to get a closer approximation to the asymptotic periodic swimming motion.
The final period in the numerical calculation is used to get the mean swimming velocity, the mean rate of dissipation, and the efficiency $E_T=\eta a^2\omega|\overline{U_{sw}}|/\overline{\mathcal{D}}$ for that period. We choose a chain with the same parameters as before, amplitude factor $\varepsilon=2$, and square scale number $s^2=10$. After fifty periods we find
 \begin{equation}
\label{5.13}\hat{U}_{EC}=0.00105,\qquad\hat{\mathcal{D}}_{EC}=34.210,\qquad E_T=308\times 10^{-7}.
\end{equation}
After $100$ periods the efficiency has increased to $E_T=498\times 10^{-7}$. After $150$ periods it has increased to $E_T=615\times 10^{-7}$. After $200$ periods the values are
 \begin{equation}
\label{5.14}\hat{U}_{EC}=0.00235,\qquad\hat{\mathcal{D}}_{EC}=34.210,\qquad E_T=688\times 10^{-7}.
\end{equation}
It seems clear that in this manner the motion eventually becomes periodic and the values tend to the asymptotic values for the given set of actuating forces.

\section{\label{VI}Discussion}

The added mass hydrodynamic interactions, derived in Sec. II, in conjunction with similar frictional hydrodynamic interactions derived earlier \cite{5}, allow discussion of the swimming of linear chain structures of the cargo type, where one of the spheres is much larger than the others. Such structures provide a fairly realistic model of organisms with a large head or of bodies driven by small appendages. In our applications we have considered only longitudinal motions along the axis of a linear chain, but the derived hydrodynamic interactions can also be used in the description of swimming by means of a transverse wave along the chain or by wave-type motion on more general structures.

In our application to the swimming of a linear chain we study the whole range of friction and inertia, including the inertial regime, where it is dominated by added mass effects. The theory is based on approximate equations of motion of limited validity, since the hydrodynamic interactions are assumed to be instantaneous and memory effects are neglected. In addition, the phenomenon of vortex shedding is absent from the theory. Nonetheless the theory provides valuable insight into the effect of added mass on swimming.

The bilinear theory of Sec. IV has the merit that it allows determination of the optimal stroke, at least for small amplitude swimming. The discussion shows that in the present model the bilinear theory provides accurate results over a wide range. This suggests that the first two terms in an expansion of the flow in the amplitude of stroke may yield valuable information. In earlier work about inertial effects in swimming \cite{13}\cite{15} we have discussed the contribution of Reynolds stress to the swimming velocity. It would be worthwhile to investigate how to incorporate the effect in the theory discussed above.

In Sec. V we have shown that for chains with a cargo constraint the optimal actuating forces can be found from a reduced bilinear theory. We showed also that the approximate equations of motion can be solved numerically to study transient effects in swimming.

\newpage

\section*{Figure captions}

\subsection*{Fig. 1}
Schematic shape of longitudinal swimmer consisting of three beads and a cargo sphere.

\subsection*{Fig. 2}
Plot of the trajectory of the four spheres along the $x$-axis in the center frame during one period in the motion given by Eqs. (3.5) and (3.7) for equilibrium distances $(d,d,b+d)$ for $d=5a$ and $b=10a$ and amplitude factor $\varepsilon=2$.

\subsection*{Fig. 3}
Plot of the maximum eigenvalue of the eigenvalue problem Eq. (4.15) for four neutrally buoyant spheres with equilibrium distances $(d,d,b+d)$ for $d=5a$ and $b=10a$ as a function of square scale number $s^2$, defined in Eq. (3.3).

\newpage
\setlength{\unitlength}{1cm}
\begin{figure}
 \includegraphics{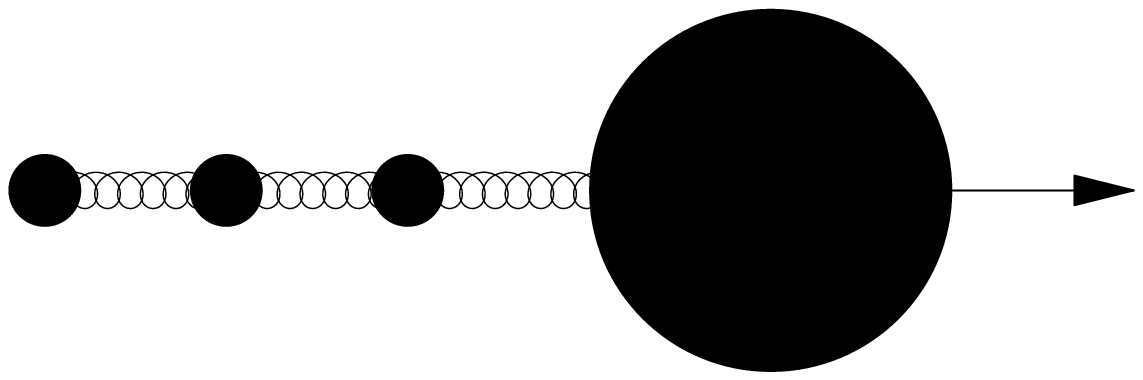}
   \put(-9.1,3.1){}
\put(-1.2,-.2){}
  \caption{}
\end{figure}
\newpage
\clearpage
\newpage
\setlength{\unitlength}{1cm}
\begin{figure}
 \includegraphics{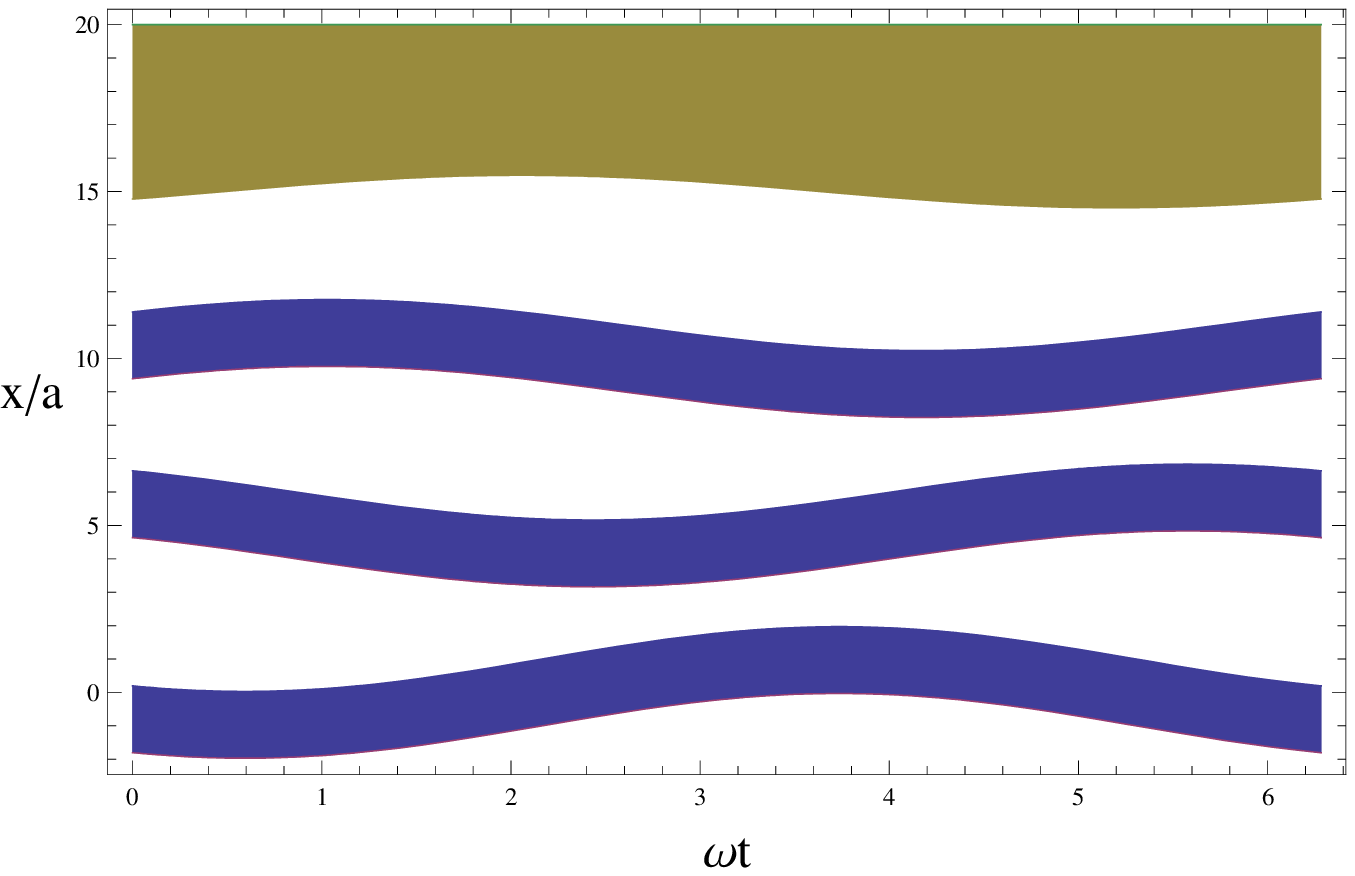}
   \put(-9.1,3.1){}
\put(-1.2,-.2){}
  \caption{}
\end{figure}
\newpage
\clearpage
\newpage
\setlength{\unitlength}{1cm}
\begin{figure}
 \includegraphics{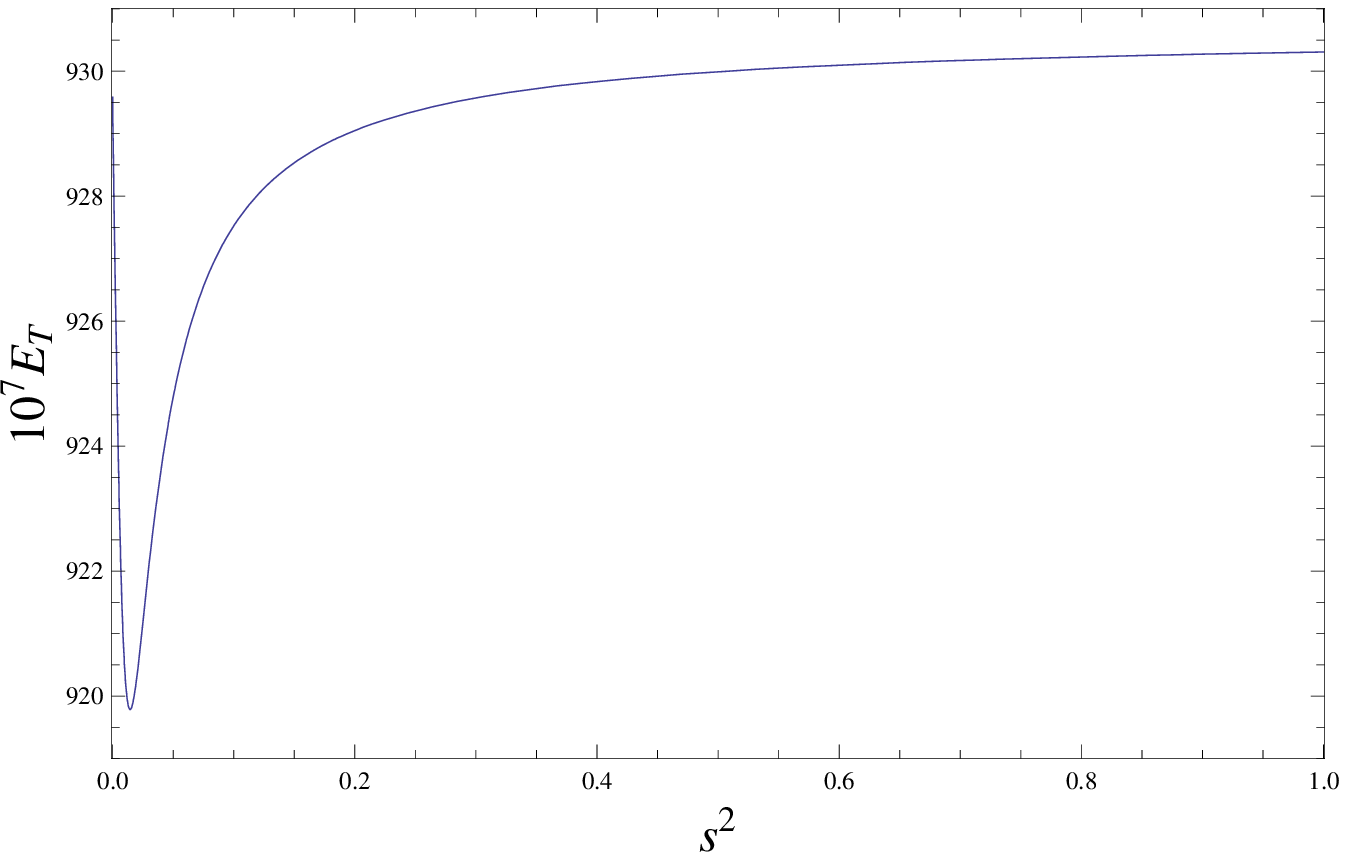}
   \put(-9.1,3.1){}
\put(-1.2,-.2){}
  \caption{}
\end{figure}
\newpage
\clearpage

\end{document}